\documentclass[11pt]{article}
\usepackage{fullpage} 
\usepackage[margin=2.5cm,top=2cm,bottom=3cm,centering]{geometry}
\usepackage{graphicx}
\usepackage[version=3]{mhchem}
\usepackage{amsmath} 
\usepackage{amsfonts}
\usepackage{multicol}

\title{Simply conceiving the Arrhenius law and absolute kinetic constants using the geometric distribution}

\author{Denis Michel\\
\\
       Universite de Rennes1-IRSET. Campus de Beaulieu Bat. 13.\\
        35042 Rennes cedex. \ denis.michel@live.fr}

\date{} 

\begin{document}
\maketitle
\begin{multicols}{2}
\textbf{Although first-order rate constants are basic ingredients of physical chemistry, biochemistry and systems modeling, their innermost nature is derived from complex physical chemistry mechanisms. The present study suggests that equivalent conclusions can be more straightly obtained from simple statistics. The different facets of kinetic constants are first classified and clarified with respect to time and energy and the equivalences between traditional flux rate and modern probabilistic modeling are summarized. Then, a naive but rigorous approach is proposed to concretely perceive how the Arrhenius law naturally emerges from the geometric distribution. It appears that (1) the distribution in time of chemical events as well as (2) their mean frequency, are both dictated by randomness only and as such, are accurately described by time-based and spatial exponential processes respectively.}
\newline
\section{Introduction}
Over zero kelvin, matter particles move, collide and react with a certain probability. All these events result from a transfer of energy provided by either radiation or previous motions and ultimately from a strong initial impulsion spreading in the universe. Dynamic modeling of all molecular systems, from enzymology to systems biology, is grounded on the generalized use of kinetic rates, usually written "\textit{k}" in traditional biochemistry. Molecular events cover a wide range of phenomena including motion, covalent modifications or interactions. Although they are associated to all these events, kinetic rates are subtle enough. Their physical meaning is generally ignored by biochemists, who need only using the famous equation of Arrhenius \cite{Arrhenius} as a general relationship for the temperature dependence of rates of reaction. According to this equation, derived from the thermodynamic study of van't Hoff \cite{Van't Hoff}, the logarithm of the rate constant is negatively proportional to the inverse of temperature. Beside this macroscopic thermodynamic approach, microscopic rate theories have then been developed, for example based on the probability of occurrence of a transition state theory \cite{Eyring} or the diffusion out of an energy well \cite{Kramers}. For complete reviews on the subject, see \cite{Hanggi,Pollak,Zhou}. Very atomistic models will not be considered here but the simplified version of the transition state theory most widespread in biochemistry textbooks will be shown to suffer from several drawbacks. The present study is aimed at recovering formally the Arrhenius law and establishing a simple definition of kinetic constants, expurgated from mechanistic considerations, using minimalist hypotheses and simple statistical tools.

\section{Modeling the events}
\subsection{The classical "mass action" approach to molecular events.}
The very founder definition of chemical transitions is based on rate fluxes, which are assumed to be proportional to the amounts of transformable reactants. In this view, which long proved very efficient, the constants \textit{k} are coefficients of proportionality whose dimensions allow to equalize the units in kinetic equations ($ \textup{time}^{-1} $ for first order reactions, or $ \textup{M}^{-1} \textup{time}^{-1} $ for second order reactions). For example, the elementary transition $ S \overset{k}{\rightarrow}P $ can be modeled as follows

\begin{subequations} \label{E:gp}
\begin{equation} -\frac{d[S]}{dt}=k[S] \end{equation} \label{E:gp1}
or
\begin{equation} \frac{d[S]}{[S]}=-kdt \end{equation} \label{E:gp2}
yielding after integration,
\begin{equation} \ln \frac{[S]_{t}}{[S]_{0}}=-k(t-t_{0})\end{equation} \label{E:gp3}
or
\begin{equation} [S]_{t}=[S]_{0} \textup{e}^{-kt} \end{equation} \label{E:gp4}
Reciprocally,
\begin{equation} [P]_{t}=[S]_{0}-[S]_{t}=[S]_{0}(1- \textup{e}^{-kt})\end{equation} \label{E:gp5}
\end{subequations}

The probabilistic view described below can be anticipated from this classical mass action approach if identifying the proportions of molecules with probabilities, but the probabilistic view can be introduced even more straightly.

\subsection{Minimalist mathematical approaches more suitable for single events in biochemistry}

A living cell is not a dish as those used in chemistry labs, because certain subcellular compartments contain very few macromolecules of the same kind. For example, the notion of concentration is obviously meaningless for a gene from the X chromosome, present in a single copy in each cell. To circumvent this problem, the single molecule mathematical approach is preferable, illuminating for understanding biochemical mechanisms \cite{Ninio}, and now widely used for modeling the nonlinear mechanisms responsible for the refined behaviors of living systems \cite{Michel2011}. The basic condition necessary to understand molecular events is that they occur randomly, that is to say without memory. If a given event is memoryless and has a mean waiting time $ \left \langle T \right \rangle $, then, the only possible law governing its probability of occurrence is the exponential distribution. At every time point \textit{t}, this probability is simply

\begin{subequations} \label{E:gp}
\begin{equation} P(X > t)= \textup{e}^{-\frac{t}{\left \langle T \right \rangle}} \end{equation} \label{E:gp1}
\begin{equation} P(X \leq t)= 1- \textup{e}^{-\frac{t}{\left \langle T \right \rangle}} \end{equation} \label{E:gp2}
\end{subequations} 

This result can be easily demonstrated. The absence of memory can be explicitly transcribed into\\
$ \forall t_{1},t_{2} \in \mathbb{R}_{+} $
\begin{equation} P(X>t_{1}+t_{2}\cap X>t_{1})=P(X>t_{2}) \end{equation}
Indeed, $ P(X>t_1+t_2|X>t_1)= P({X>t_1+t_2} \cap {X>t_1} )/P(X>t_1)
=P(X>t_1+t_2)/P(X>t_1) $\\

Hence, Eq.(3) becomes

\begin{subequations} \label{E:gp}
\begin{equation} P(X>t_{1}+t_{2})=P(X>t_{1})P(X>t_{2}) \end{equation}\label{E:gp1}

This relationship imposes to handle \textit{t} as an exponent, such that

\begin{equation} \textup{e}^{-k(t_{1}+t_{2})}=\textup{e}^{-kt_{1}}\textup{e}^{-kt_{2}} \end{equation}\label{E:gp2}
\end{subequations} 

in which \textit{k} is introduced to cancel the dimension of the exponent and turns to correspond to a frequency ($ \textup{time}^{-1} $) and the average transition time is
\begin{equation} \left \langle T \right \rangle =\int_{t=0}^{\infty } kt\ \textup{e}^{-kt} dt=\frac{1}{k} \end{equation} 

\subsection{The ergodic correspondence}
The recent outset of single molecule-based perspectives, particularly in biochemistry, radically changed the traditional concepts of chemistry (Table 1). 

\end{multicols}
\begin{table}[!h]
\caption{Correspondences between the traditional bulk approach of chemistry and single molecule-based perspective}
\label{tab:1}       
\begin{tabular}{ll}
\hline\noalign{\smallskip}
Mass action approach & Single molecule approach  \\
\noalign{\smallskip}\hline\noalign{\smallskip}
Concentration & State probability  \\
Rate flux & Poisson parameter  \\
Total concentration of the leading macromolecule & 1 
(so that $ V_{max}=k_{cat} $ in enzymology)  \\
Space & Time  \\
\noalign{\smallskip}\hline
\end{tabular}
\end{table}
\begin{multicols}{2}
Of particular interest is the replacement of space (in which the concentration of reactants is defined), by time. This switch is fundamentally rooted in the principle of ergodicity, according to which the collection of states taken at a single time by an infinite number of particles, corresponds to those taken by a single particle during an infinite time window. Experimental studies suggest that this assumption is acceptable, as illustrated by the single enzyme activity monitored in \cite{Lu}. At equilibrium, the classical Michaelis-Menten fraction of occupied macromolecules can be defined as well as a fractional occupation time of a single macromolecule \cite{Michel}. Although the principle of ergodicity is generally presented as a non-demonstrated axiom, the identical results obtained from the mass action (section 2.1) and mathematical (section 2.2) approaches allow to understand its basis.

\section{Modeling the frequencies of the events}
\subsection{Inconsistensies in the classical description of the transition state theory}
The version of the transition state theory most widespread in textbooks suffers from several drawbacks. Transitions from a starting point \textit{S} to a final point \textit{P} are supposed to follow a certain delay because they are restricted by an energy barrier. The transition is conditioned by preliminary passage through an unstable reaction intermediate named activated complex and usually written with a double-dagger label $ \ddagger $. The resulting two-step reaction is represented in Fig.1.

\begin{center}
\includegraphics[width=5cm]{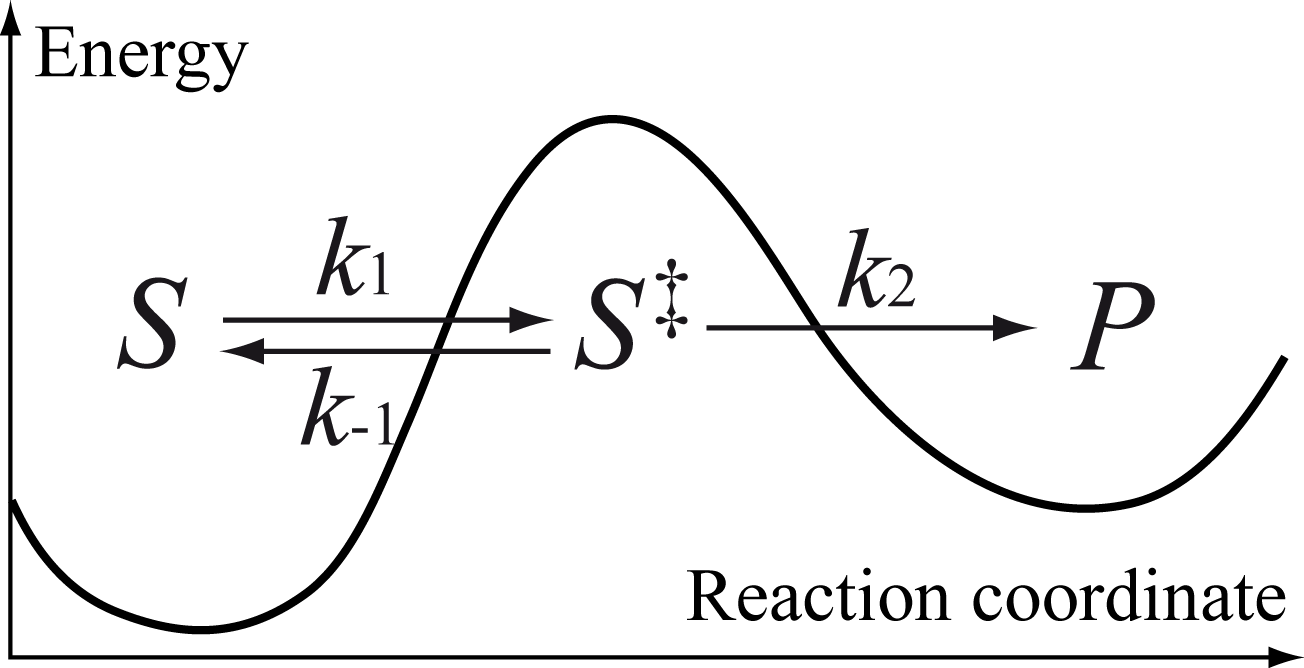} \\
\end{center}
\begin{small} \textbf{Fig.1.} The traditional energy barrier responsible for reaction delays.\end{small}\\
\newline
It is intriguing that when using this reasoning, the elementary rate \textit{k} to be defined, is now replaced by 3 such rates ($ k_1 $, $ k_{-1} $ and $ k_2 $). The $ S^{\ddagger } $ intermediate is considered as short-lived. Manipulating rates as parameters of exponential functions implies that the global transition from \textit{S} to \textit{P} takes the form $ k=k_1 k_2/(k_{-1}+k_2) $. New hypotheses should be introduced to simplify this value

\textbf{1st hypothesis, $ k_2 >> k_{-1} $}. Once generated, $ S^{\ddagger } $ rapidly converts into \textit{P} because of the absence of energetic barrier between $ S^{\ddagger } $  and \textit{P}. One haves $ k=k_1 $ so that the definition of \textit{k} is simply replaced by that of $ k_1 $, no way solving the initial question: what is a \textit{k}?

\textbf{2d hypothesis, $ k_2 << k_{-1} $}. This is the hypothesis classically retained in the activated complex theory. In this case, $ k=k_2(k_1/k_{-1}) $, or $ k=k_2 K^{\ddagger } $. Now, the rate constant is defined by an equilibrium (as initially proposed by Arrhenius himself, between normal and hypothetical reaction-prone molecules). 

\begin{equation} K^{\ddagger }=\frac{[S^{\ddagger }]}{[S_{0}]}=\textup{e}^{-\frac{\Delta E^ \ddagger}{k_{B}T}} \end{equation}

from which the Arrhenius law can be recovered

\begin{equation}k= A\ \textup{e}^{-\frac{E_{a}}{k_{B}T}} \end{equation}

In this expression, the mysterious constant \textit{"A"}, corresponding to a constant of integration in the thermodynamic approach, is called "pre-exponential coefficient" and is extremely important since it gives to \textit{k} the dimension of a rate ($ \textup{time}^{-1} $) \cite{Laidler}. In the present description, it corresponds to $ k_2 $. According to physical principles, this rate (written below $ k^{\ddagger } $) includes an universal component $ k_B T/h $ time$ ^{-1} $, corresponding to the mean vibrational or translational energy ($ k_B T $) where $ T $ is the temperature and $ k_B $ is Boltzmann's constant, divided by Planck's constant \textit{h}. With $ k_B = 1.38 \ 10^{-23} $ J/K, $ T $ = 300 K and $ h=6.63 \ 10^{-34} $ J.s, one obtains $ k_B T/h = 6.25 \ 10^{12} $ /s, which is very high indeed, but renders absurd the initial hypothesis  $ k_2 << k_{-1} $.

\textbf{3d hypothesis, $ k_2=k_{-1}=k^{\ddagger }$}. In principle, $ k^{\ddagger }$ is expected to be similar for the two barrier-free transitions starting from $ S^{\ddagger }$, and in the present case for the backward transition $ k_{-1} $, such that

\begin{center}
\includegraphics[width=4cm]{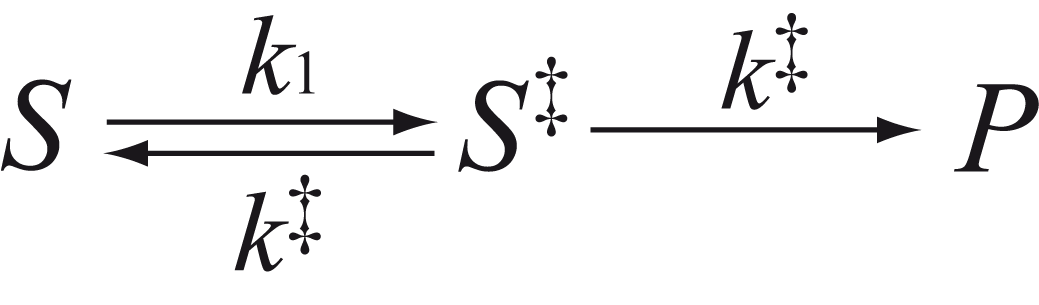} \\
\end{center}

Then, $ k=k_1 k^{\ddagger }/(k^{\ddagger } + k^{\ddagger }) = k_1 /2 $, letting unchanged the initial question. In conclusion, we see that this widespread description of rate constants in textbooks is clearly not satisfying. An alternative simple view of statistical distribution of energy can be proposed.

\subsection{The geometric approach to energy distribution for systems capable of unlimitedly storing energy}

Consider a closed system in which every particle can store energy under different forms (rotational, vibrational or translational) which all correspond to interchangeable energy quanta. If a given number of energy quanta distribute randomly over a given number of particles, then the system can be compared to a very large pearl necklace made of two kinds of pearls: \textit{B} black and \textit{W} white pearls. The white beads are supposed to correspond to energy quanta and the black beads represent the separations between the particles. Adjacent white beads clustered between two black beads correspond to the number of energy quanta (written $ \varepsilon $) in a single particle. If the beads added to the string are drawn randomly from a bag containing a huge number of well-mixed black and white pearls (equivalent to a drawing with replacement), then the probability that any given bead on the string is white is $ W/(W+B) $. Hence, the probability that an arbitrarily chosen stretch of white beads contains more than $ \varepsilon ^{\ddagger} $ beads is

\begin{equation} P(\varepsilon >\varepsilon ^{\ddagger })=\left (\frac{W}{B+W}  \right )^{\varepsilon ^{\ddagger }} \end{equation}

that can be rewritten, if defining the ratio of energy over particles $ W/B = \left \langle \varepsilon  \right \rangle $,

\begin{equation} P(\varepsilon > \varepsilon ^{\ddagger })=\left (\frac{\left \langle \varepsilon  \right \rangle}{1+\left \langle \varepsilon  \right \rangle}  \right )^{\varepsilon ^{\ddagger }} \end{equation}

For $ \left \langle \varepsilon \right \rangle $ and $ \varepsilon ^{\ddagger} $ large and of the same order of magnitude and using the property of the exponential function

\begin{equation} \lim\limits_{n \to \infty} \left (1+\frac{x}{n}  \right )^{n}=\textup{e}^{x} \end{equation}

Eq.(9) approaches 

\begin{equation} P(\varepsilon > \varepsilon ^{\ddagger })\sim  \textup{e}^{-\frac{\varepsilon ^{\ddagger }}{\left \langle \varepsilon  \right \rangle}} \end{equation}

This exponential result is naturally expected for large numbers of beads. Indeed, given that the probability that any given bead on the string is black is $ B/(W+B) $, then, when starting from a black bead and walking along the string, the number of beads to be examined to find the following black one is, on average, the reciprocal of the previous value  $ (W+B)/B $. As a consequence, the average number of white beads in each interval is  $ [(W+B)/B] - 1 = W/B $, which is the previously defined entity $ \left \langle \varepsilon  \right \rangle $, or mean number of energy quanta per particle. When switching to a spatially continuous perspective by considering a large pearl necklace with infinitesimally small beads, $ \varepsilon  $ is the length of a white segment. Given the randomness of bead sequence, the probability that a segment is longer than $ \varepsilon ^{\ddagger} $ necessarily follows an exponential distribution such that

\begin{equation} P(\varepsilon > \varepsilon ^{\ddagger })= \textup{e}^{-\frac{\varepsilon ^{\ddagger }}{\left \langle \varepsilon  \right \rangle}} \end{equation}

According to this probability, the kinetic rate would be simply
\begin{equation} k=k ^{\ddagger } \textup{e}^{-\frac{\varepsilon ^{\ddagger }}{\left \langle \varepsilon  \right \rangle}} \end{equation}
It is of course tempting to connect this result to Boltzmann by identifying $ \left \langle \varepsilon  \right \rangle $ with the mean energy per particle. This exponential function should not be confused with the one defined in section 2.2, in which the unidimensional intensity was conceived along time, whereas it is now specified along the energy content. Consequently, the probability that an event occurs at every time point is an exponential of exponential
\begin{equation} P(X > t)= \textup{e}^{-k ^{\ddagger }t. \textup{e}^{-\frac{\varepsilon ^{\ddagger }}{\left \langle \varepsilon  \right \rangle}}} \end{equation}

This simple building of the Arrhenius law is in fact related to the Bose-Einstein distribution, assuming that energy quanta are inherently immaterial, permutable between the particles and indefinitely superposable in a single particle. According to the exponential relaxation observed for many phenomena, this distribution is known to approach the Maxwell-Boltzmann distribution at high temperature. Indeed, the probability that a white stretch contains precisely $ \varepsilon ^{\ddagger} $ beads follows a geometric law plus 1, such that the $ (\varepsilon ^{\ddagger} +1) $th bead to be drawn should be black. Hence,
\begin{equation} P(\varepsilon = \varepsilon ^{\ddagger })=\left (\frac{\left \langle \varepsilon  \right \rangle}{1+\left \langle \varepsilon  \right \rangle}  \right )^{\varepsilon ^{\ddagger }}\frac{1}{1+\left \langle \varepsilon  \right \rangle} \end{equation}

illustrated in Fig.2 and whose maxima are precisely obtained when $ \left \langle \varepsilon \right \rangle =\varepsilon ^{\ddagger} $. This simple probability distribution indeed approaches the Boltzmann's one when $ \varepsilon $ is large enough. Eq.(15) can be converted into its continuous version
\begin{equation} P(\varepsilon = \varepsilon ^{\ddagger })\sim  \textup{e}^{-\frac{\varepsilon ^{\ddagger }}{\left \langle \varepsilon  \right \rangle}}\left (1- \textup{e}^{-\frac{1}{\left \langle \varepsilon  \right \rangle}}  \right ) \end{equation}
\begin{center}
\includegraphics[width=6cm]{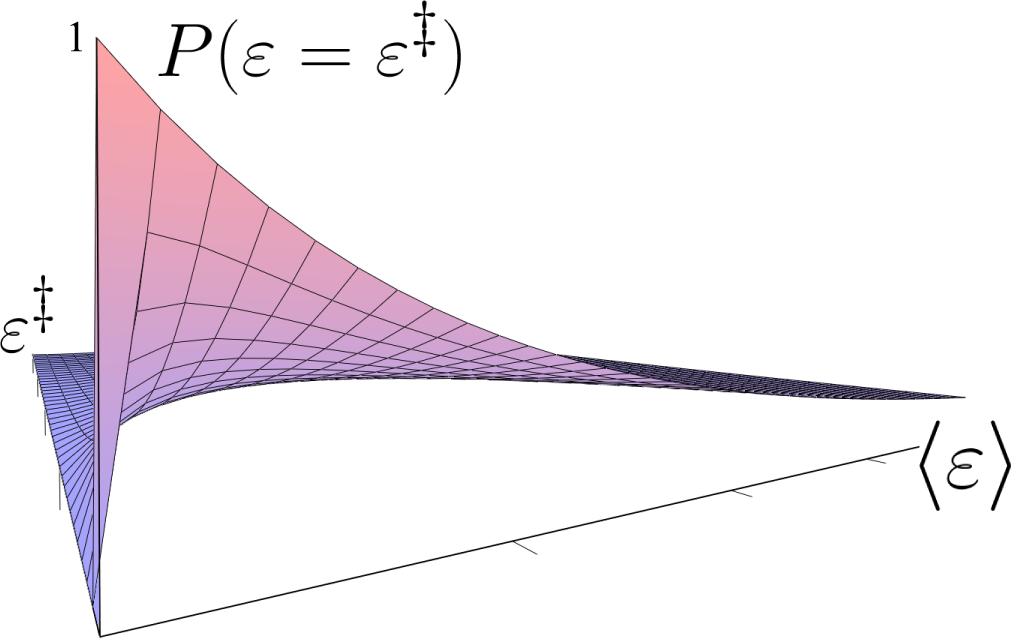} \\
\end{center}
\begin{small} \textbf{Fig.2.} Probability distribution of particles with energy $ \varepsilon ^{\ddagger } $, when increasing the mean energy per particle $ \left \langle \varepsilon  \right \rangle  $ (that is to say the temperature).\end{small}\\

and given that

\begin{equation} \sum_{j =0}^{\infty }\textup{e}^{-\frac{j}{\left \langle \varepsilon  \right \rangle}}=\left (1- \textup{e}^{-\frac{1}{\left \langle \varepsilon  \right \rangle}}  \right )^{-1} \end{equation}

one obtains 

\begin{equation} P(\varepsilon = \varepsilon ^{\ddagger })\sim \frac{ \textup{e}^{-\frac{\varepsilon ^{\ddagger }}{\left \langle \varepsilon  \right \rangle}}}{\sum_{j =0}^{\infty }\textup{e}^{-\frac{j}{\left \langle \varepsilon  \right \rangle}}} \end{equation}

which is similar to the celebrated formula of Boltzmann, obtained using the entropy maximizing procedure and which predicts that the probability that any particle has an amount of energy $ E_{i} $ is 

\begin{equation} P(E_{i})=\textup{e}^{-\beta E_{i}}/ \sum _{j=0}^{n}  \textup{e}^{-\beta E_{j}} \end{equation}

where $ \beta = 1/k_B T $ and the denominator is the canonical partition function summed over the \textit{n} possible microstates \cite{Pathria}. Hence, there is more than an analogy between the Boltzmann and geometric treatments for the infinitesimal probabilities of large systems. For small systems, the simple geometric plus 1 law of Eq.(15) is preferable. The value of the rate constant of Eq.(13) obtained using the simple necklace toy model, is also very similar to the transition-state rate derived from a Hamiltonian, described for example in \cite{Hanggi}.\\

Most real systems are made of mixtures of molecules of different kinds, that can eventually convert into each-others, but also continuously exchange their energy quanta. These generalized exchanges have large-scale energy buffering effects allowing the mean energy per type of particle to remain roughly constant, regardless of the size of the considered subpopulation. This point is important to understand that the definition of rate constant proposed above remains true for transient fluxes in isothermal environments.

\section{Transient fluxes}
The above developments hold in equilibrium conditions, when the proportions of particles of the same kind remain constant. This situation was well described by Lewis \cite{Lewis} who compared the different populations of molecules to the habitants of different towns. Although people are constantly moving between different towns, the relative town sizes are metastable and remain roughly constant in equilibrium. But imagine now that a given node is pulled out from the network, such that the departures of molecules are no longer compensated by arrivals: Is it possible to determine the exit rate in this situation? Different cases will be examined, under the assumption that the exchanges of energy quanta in the residual population of particles are more rapid than the escape of particles.

\subsection{Particle efflux in isothermal conditions}
This is the realistic situation holding for transient phenomena in living cells, in which the overall amount of energy quanta over all molecular components roughly remains constant. Under this condition, although the arrival of new particles is precluded, the mean energy per resident particle $ \left \langle \varepsilon  \right \rangle $ remains constant, owing to collisions with other types of components. In this case, the node is expected to lose all its components with the rate defined in Eq.(13), in the same manner that all boiling water disappears from a ban that is constantly heated.

\subsection{Efflux of particles from an energetically isolated node}
If, in addition to particle arrivals, the import and export of pure energy between the node and its environment are also precluded, then, the efflux rate obviously decreases with time since energy quanta accompany escaped particles. Moreover, energy quanta are expected to disappear faster than particles since only the most energy-rich particles are filtered by the threshold $ \varepsilon ^{\ddagger } $ for flying away. Interestingly, the drop in total energy resulting from the evasion of the first particles can prevent the remaining particles to obtain their own travelling passport, rendering them definitely prisoners in absence of energy refueling. A good illustration of this situation is the rate of evaporation of water molecules from a hot soup. Ordinary experience shows that evaporation decreases as the soup cools. For modeling this, let us transform the total number of energy quanta in the soup by a continuous variable $ x(t) $ decreasing with time and the total number of particles by a continuous variable $ y(t) $ also decreasing with time. Let us suppose that the medium is not sticky to allow the particles to clear off once their energy content reaches $ \varepsilon ^{\ddagger } $. Then, the efflux rate derived from the geometric distribution depends on the evolution of $ x(t) $ and $ y(t) $ and obey the following system 

\begin{subequations} \label{E:gp}
\begin{equation} k(t)=k^{\ddagger } \left (\frac{x(t)}{y(t)+x(t)}  \right )^{\varepsilon ^{\ddagger }} \end{equation} \label{E:gp1}
with
\begin{equation}  \dot{x}(t)=\varepsilon ^{\ddagger } \dot{y}(t) \end{equation} \label{E:gp2}
and
\begin{equation} \dot{y}(t)= -k(t)y(t) \end{equation} \label{E:gp3}
\end{subequations} 

The fate of this system (Fig.3), depends on the relative values of the reaction threshold $ \varepsilon ^{\ddagger } $ and of the starting amounts $ x_{0} $ and $ y_{0} $. If $ y_{0} \leq  x_{0}/ \varepsilon ^{\ddagger } $, then all the particles disappear, but if $ y(t_{0}) > x_{0}/ \varepsilon ^{\ddagger } $, then a fraction of the initial population of particles ($ y_{0} - x_{0}/ \varepsilon ^{\ddagger } $) remains stable, such that

\begin{equation} y_{\infty }= y_{0}  \left ( 1-\frac{\left \langle \varepsilon  \right \rangle _{0}}{\varepsilon ^{\ddagger }} \right ) \end{equation} 

It is amusing, but of course not necessary, to recourse to Bose-Einstein to conceive that the escape of the highest energy water molecules in the form of vapor, can rapidly cool a bowl of hot soup without significantly reducing its volume, in line with the right panel of Fig.3. It is also clear that stirring the soup accelerates its cooling by allowing the molecules whose energy content is sufficient, to fly away instead of unnecessarily accumulating more energy.

\end{multicols}
\begin{center}
\includegraphics[width=12cm]{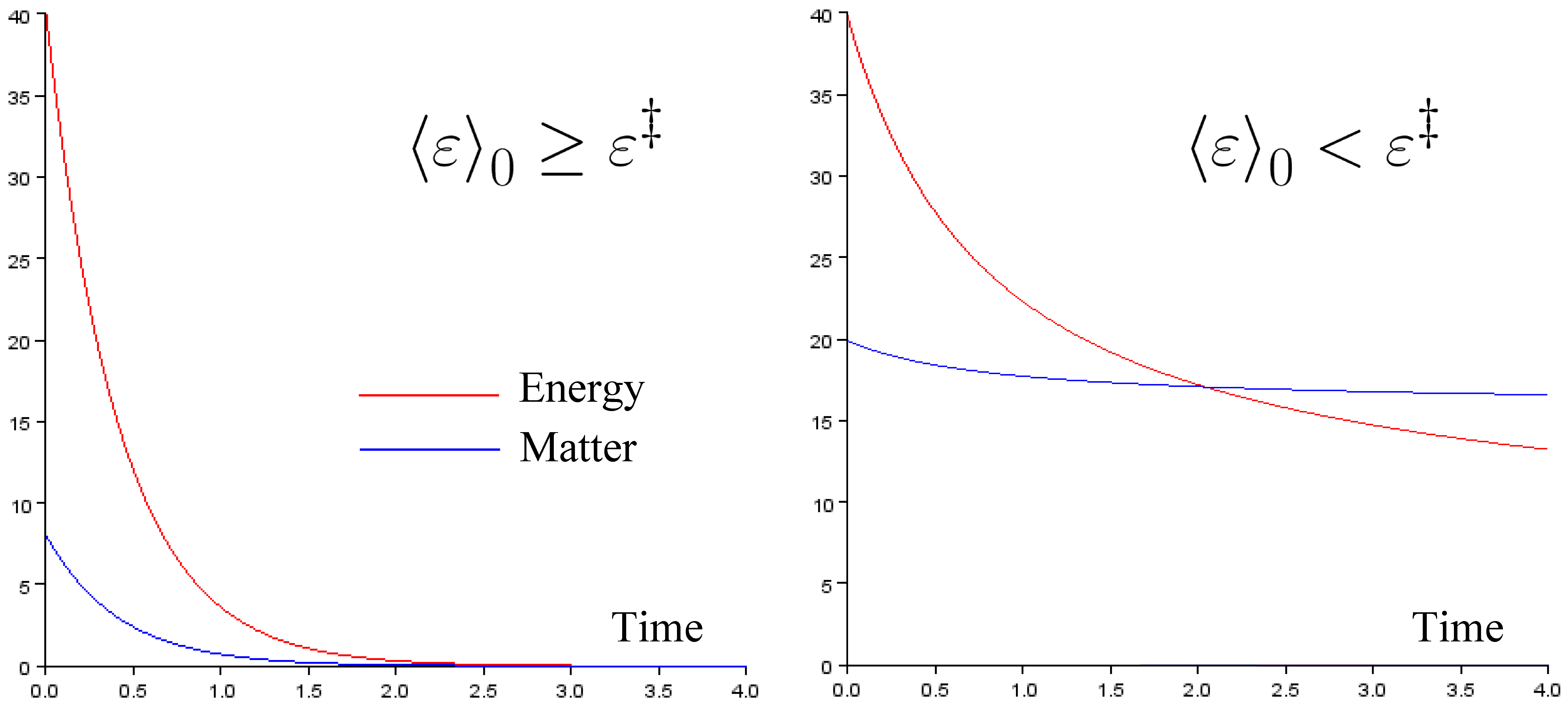} \\
\end{center}
\begin{small} \textbf{Figure 3.} Transient escape of energy (red lines) and particles (dark blue lines) from an energetically-isolated subsystem. This unstable system can lose all its components or not, depending on the ratio between the initial mean energy per particle and the threshold transition energy. Curves drawn to Eq.(20) using $ (k^{\ddagger }, \varepsilon ^{\ddagger }, x_{0}, y_{0} $) = (6, 5, 40, 8) and (6, 8, 40, 20) for the left and right panel respectively. \end{small}\\
\begin{multicols}{2}

\section{Discussion}
It is suggested here that the geometric distribution is a simple substitute to traditional approaches to easily conceive the Arrhenius law, without need for the accessory tools of statistical mechanics such as the Stirling approximation and Lagrange multipliers and without recourse to the particular mechanisms currently used in rate theories, including collision, diffusion or S-matrix theories. The fact that equivalent resulting behaviors are more directly obtained through conveniently selected statistical approaches is not a surprise. Boltzmann long proved that statistical and enumerating strategies are shortcuts to recover previous physical results. Moreover, the quest for simplicity does not forbid rigor, according to the pragmatic recommendation of Josiah Willard Gibbs: "\textit{One of the principal objects of theoretical research in any department of knowledge is to find the point of view from which the subject appears in its greatest simplicity}". This reductionist study suggests that the distribution in time of chemical events, as well as the energetic threshold over which they occur, are both fundamentally related to the exponential law, that is itself, together with its discrete counterpart the geometric law, the exact formulation of randomness.\\
\newline
\textbf{Ackowledgement} The author thanks Jean-Christophe Breton and Benjamin Boutin for helpful discussions.

\end{multicols}
\end{document}